\documentclass[12pt,letter]{article}
\pdfoutput=1
\usepackage{graphicx, epsfig, color,cite}
\usepackage{amsmath}
\usepackage{amssymb}
\usepackage{float}
\usepackage{caption,subcaption,graphicx}
\usepackage{hyperref}

\def\bi{\begin{itemize}}
\def\ei{\end{itemize}}

\def\tst{\tilde t}

\def\alt{\lesssim}
\def\agt{\gtrsim}
\def\be{\begin{equation}}  
\def\ee{\end{equation}}  
\def\bea{\begin{eqnarray}}  
\def\eea{\end{eqnarray}}

\begin{document}
\begin{titlepage}
\begin{flushright}
OU-HEP-220204
\end{flushright}

\vspace{0.5cm}
\begin{center}
  {\Large \bf Radiative natural supersymmetry \\
    emergent from the string landscape}\\
\vspace{1.2cm} \renewcommand{\thefootnote}{\fnsymbol{footnote}}
{\large Howard Baer$^{1,2}$\footnote[1]{Email: baer@ou.edu },
Vernon Barger$^2$\footnote[2]{Email: barger@pheno.wisc.edu},
Dakotah Martinez$^1$\footnote[3]{Email: dakotah.s.martinez-1@ou.edu} and
Shadman Salam$^1$\footnote[3]{Email: shadman.salam@ou.edu}
}\\ 
\vspace{1.2cm} \renewcommand{\thefootnote}{\arabic{footnote}}
{\it 
$^1$Homer L. Dodge Department of Physics and Astronomy,
University of Oklahoma, Norman, OK 73019, USA \\[3pt]
}
{\it 
$^2$Department of Physics,
University of Wisconsin, Madison, WI 53706 USA \\[3pt]
}

\end{center}

\vspace{0.5cm}
\begin{abstract}
\noindent
In string theory with flux compactifications, anthropic selection
for structure formation from a discretuum of vacuum energy values 
provides at present our only understanding of the tiny yet positive 
value of the cosmological constant. 
We apply similar reasoning to a toy model of the multiverse restricted to
vacua with the MSSM as the low energy effective theory.
Here, one expects a statistical selection favoring large soft SUSY 
breaking terms leading to a derived value of the weak scale
in each pocket universe (with appropriate electroweak symmetry breaking)
which differs from the weak scale as measured in our universe. 
In contrast, the SUSY preserving $\mu$ parameter is selected uniformly 
on a log scale as is consistent with the distribution of SM fermion masses:
this favors smaller values of $\mu$.
An anthropic selection of the weak scale to within a factor of a few of 
our measured value-- in order to produce complex nuclei as we know them 
(atomic principle)-- provides statistical predictions for Higgs and 
sparticle masses in accord with LHC measurements. 
The statistical selection then more often leads to
(radiatively-driven) {\it natural} SUSY models 
over the Standard Model or finely-tuned SUSY models
such as mSUGRA/CMSSM, split, mini-split, spread, high scale or PeV SUSY.
The predicted Higgs and superparticle spectra might be testable 
at HL-LHC or ILC via higgsino pair production but is certainly testable at 
higher energy hadron colliders with $\sqrt{s}\sim 30-100$ TeV.
\end{abstract}
\end{titlepage}

\section{Introduction}
\label{sec:intro}

How can it be that the vacuum energy density 
$\rho_{vac}=\Lambda_{cc} c^2/8\pi G_N=\Lambda_{cc}m_P^2\sim (0.003\ {\rm eV})^4$ 
is more than 120 orders of magnitude below its expected value 
from quantum gravity? Weinberg suggested that in an eternally inflating 
multiverse\cite{Guth:2000ka,Linde:2015edk} with each {\it pocket universe} (PU) supporting its own non-zero 
value of the cosmological constant (CC) $\Lambda_{cc}$, 
and with $\Lambda_{cc}$ being distributed across the decades 
of allowed values, the value of $\Lambda_{cc}$ ought to be
no larger than the critical value for which large scale structure, 
which is required for life as we know it to emerge. 
This allowed Weinberg to predict the value of $\Lambda_{cc}$ to a factor of 
several over a decade before it was observed\cite{Weinberg:1987dv,Martel:1997vi}.

Weinberg's prediction relied on {\it environmental selection} of a fundamental 
constant of nature. His solution to the CC problem found a home in 
a more nuanced understanding of string theory vacuum states\cite{Bousso:2000xa}. 
In compactified string theory, one expects the emergence of a visible
sector containing the Standard Model (SM) along with a variety of hidden 
sectors and a large assortment of moduli fields: gravitationally coupled 
scalar fields that determine the size and shape of the compactified 
manifold and whose vacuum expectation values determine most of the 
parameters of the $4-d$ low energy effective field theory (EFT).
In realistic flux compactifications of type IIB
string theories\cite{Douglas:2006es}, a common estimate for the number of distinct 
(metastable) vacua can range 
up to $10^{500-1000}$\cite{Ashok:2003gk}, and even more for $F$-theory compactifications\cite{Taylor:2015xtz}. These vacua, each with its own $4-d$ EFT and value for $\Lambda_{cc}$,
are more than enough to support Weinberg's solution to the CC problem.

While string theory contains only one scale, the string scale $m_s$,
Weinberg's solution provides a mechanism for the emergence of a new scale, 
$\Lambda_{cc}\ll m_s$ via environmental (or anthropic) selection.
This result obtains from the expected CC probability distribution
\be
dN_{vac}\sim f_{cc}\cdot f_{structure}\cdot d\Lambda_{cc}
\ee
where $dN_{vac}$ is the differential distribution of vacua in terms of 
the cosmological constant. 
Weinberg assumed the distribution $f_{cc}(\Lambda_{cc})$ was uniformly 
distributed in the vicinity of $10^{-120}m_P^2$. Also,
$f_{structure}(\Lambda_{cc})$ had the form of a step function
$f_{structure}\sim \Theta (10^{-120}m_P^2-\Lambda_{cc})$ such that values of
$\Lambda_{cc}$ too much bigger than our (to be) observed value would lead
to too rapid cosmological expansion so that structure in the form of
condensing galaxies (and hence stars and planets) would not form, and hence
observors would not arise.
  
Can similar reasoning be applied to the origin of other scales 
such as the weak scale? Indeed, Agrawal {\it et al.}\cite{Agrawal:1997gf,Agrawal:1998xa} (ABDS) addressed
this question in 1998. They found that-- in order to allow the formation of
complex nuclei, and hence atoms as we know them which seem essential for life
to emerge-- the allowed values of the weak
scale are located within a rather narrow window of values (the ABDS window).
Our measured value of $m_{weak}\sim m_{W,Z,h}$ seems to be centrally located 
within the ABDS window which extends roughly from $0.5 m_{weak}^{OU}- (2-5)m_{weak}^{OU}$,
where $m_{weak}^{OU}$ is the measured value of the weak scale in our universe.

In the case of the SM, with Higgs potential
$V_{Higgs}=-\mu_{SM}^2(\phi^\dagger\phi )+\lambda (\phi^\dagger\phi)^2$, with
$\lambda >0$ to ensure stability of the Higgs vev, then one might expect
\be
dN_{vac}\sim f_\mu (\mu_{SM} )\cdot f_{ABDS}(\mu_{SM} ) .
\ee
If one assumes {\it all} scales of $\mu_{SM}$ equally likely, then
$f_\mu\sim 1/\mu_{SM}$. Meanwhile, if the PU value of the weak scale
$m_{weak}^{PU}\agt (2-5)m_{weak}^{OU}$ (where $OU$ refers to the measured
value in {\it our universe}), then the up-down quark mass difference would
grow to such an extent that neutrons would no longer be stable within nuclei.
Consequently, nuclei consisting of multiple protons would no longer be stable
(too much Coulumb repulsion), and the only stable nuclei would consist of
single proton states: the universe would be chemically sterile and life
as we know it would not arise. This argument has been used as an alternative 
to the usual naturalness argument in that using anthropic reasoning, then the SM might well be valid all the way up to huge scales 
$Q\sim m_{GUT}-m_s$\cite{Elias-Miro:2011sqh} in spite of the presence of quadratic divergences in
the Higgs boson mass-squared.

Environmental selection can also be applied to supersymmetric models wherein
the Higgs mass-squared contains only logarithmic divergences. 
Indeed it is emphasized in Ref. \cite{Baer:2019cae} that in a landscape containing 
comparable numbers of SM-like and weak scale SUSY-like low energy EFTs, then
the anthropically allowed SUSY models should be much more prevalent 
because there should be a far wider range of natural parameter choices
available compared to the finetuned values which are required for the SM.
For SUSY models, we expect a distribution of soft term values
according to 
\be
dN_{vac}\sim f_{SUSY}(m_{soft})\cdot f_{EWSB}\cdot dm_{soft}
\ee 
For the soft term distribution $f_{SUSY}(m_{soft})$,
positive power
law\cite{Douglas:2004qg,Susskind:2004uv,Arkani-Hamed:2005zuc,Broeckel:2020fdz}
or log\cite{Baer:2020dri}
distributions pull soft terms to large values and seem favored by LHC 
SUSY search results\cite{Baer:2016lpj,Baer:2017uvn,Baer:2020dri,Baer:2019tee}.
In contrast, negative power law distributions, 
as expected in dynamical SUSY breaking where all SUSY breaking scales would 
be equally favored\cite{Dine:2004is,Dine:2005iw,Dine:2004ct}, or large-volume scenario (LVS)
compactifications\cite{Broeckel:2020fdz}would lead to sparticle masses
below LHC limits and light Higgs boson masses much lighter than the
measured value $m_h\sim 125$ GeV\cite{Baer:2021uxe}.

The anthropic selection function $f_{EWSB}$ requires that the derived value 
of the weak scale in each pocket universe 
\be
m_Z^{PU2}/2=\frac{m_{H_d}^2+\Sigma_d^d-(m_{H_u}^2+\Sigma_u^u)\tan^2\beta}{\tan^2\beta -1}-\mu^2
\label{eq:mzsPU}
\ee
lies within the ABDS window. Thus, one must {\it veto} MSSM-like
pocket universes wherein $m_Z^{PU}>(2-5)m_Z^{OU}$ where 
$m_Z^{OU}=91.2$ GeV is the value of $m_Z$ in our universe.
Assuming no finetuning of the values entering the right-hand-side of 
Eq. \ref{eq:mzsPU}, then conventional sparticle and Higgs mass generators such as Isajet\cite{Paige:2003mg} and others\cite{Baer:2021tta} can be used to make landscape predictions for
sparticle and Higgs boson masses. Without finetuning, then the pocket universe
value for the weak scale will typically be the maximal entry on the RHS 
of Eq. \ref{eq:mzsPU}. Then, requiring $m_Z^{PU}\alt 4m_Z^{OU}$ is the same
as requiring the electroweak naturalness measure $\Delta_{EW}\alt 30$
(where
$\Delta_{EW}\equiv |maximal\ term\ on\ RHS\ of\ Eq.~\ref{eq:mzsPU} |/m_Z^{OU2}/2$)\cite{Baer:2017uvn,Baer:2012cf}. 
Coupling the ABDS requirement with a mild log or power-law
draw to large soft terms, then the probability distributions for sparticle 
and Higgs masses can be computed. It has then been found that the light Higgs 
mass distribution $dP/dm_h$ rises to a peak at $m_h\sim 125$ GeV whilst 
sparticle masses are lifted beyond present LHC search
limits\cite{Baer:2016lpj,Baer:2017uvn,Baer:2020dri,Baer:2019tee}
(see Ref. \cite{Baer:2020kwz} for a recent review).

A drawback to the above approach is that it doesn't allow for 
accidental (finetuned) parameter values conspiring to create 
$m_Z^{PU}\ne m_Z^{OU}$ values which nonetheless end up lying 
within the ABDS window. This is because the spectrum generators
all have the measured value of $m_Z$ hardwired into their 
electroweak symmetry breaking conditions.
In the present paper, we build a toy computer code which should
provide a better simulation as to what is thought to occur within 
the multiverse in the case where a subset of vacua containing the 
minimal supersymmetric standard model (MSSM) is required to be the low 
energy EFT. This approach allows us to display whether natural SUSY models
or finetuned SUSY models are more likely to arise from the landscape.
The natural SUSY models\footnote{By natural SUSY models, we mean SUSY models wherein the soft terms are driven radiatively via RGEs to natural
  {\it weak scale values}; these models are also labelled as
  radiatively-driven natural SUSY or
  radiative natural SUSY (RNS)\cite{Baer:2012up,Baer:2012cf}. The RNS SUSY models are distinct from other
  versions of natural SUSY which may require sub-TeV top squarks or sparticles
  at or around the weak scale $\sim 100$ GeV. For a distinction between models,
see {\it e.g.} Ref's \cite{Baer:2013gva,Baer:2014ica}.} are characterized by low $\Delta_{EW}\alt 30$
while finetuned SUSY models include the constrained MSSM
(CMSSM or mSUGRA model)\cite{Kane:1993td}, minisplit SUSY\cite{Arvanitaki:2012ps,Arkani-Hamed:2012fhg}, PeV SUSY\cite{Wells:2004di},
high scale SUSY\cite{Giudice:2011cg}, spread SUSY\cite{Hall:2011jd} and the G2MSSM\cite{Acharya:2008zi}.

In Sec. \ref{sec:nuhm}, we discuss our assumed SUSY model and low energy EFT
framework.
In Sec. \ref{sec:soft}, we review expectations for the landscape
soft term distribution $f_{SUSY}(m_{soft})$ and why it favors large over
small soft SUSY breaking terms. We also discuss the assumed distribution
for the SUSY $\mu$ parameter.
In Sec. \ref{sec:ABDS}, we review the crucial anthropic condition that
surviving SUSY models lie within the ABDS window, {\it i.e.} that the
pocket universe value of the weak scale is not-too-far displaced from  the
value of $m_{weak}$ in our universe. In Sec. \ref{sec:model}, we describe our
toy model simulation of the multiverse with varying values of $m_{weak}^{PU}$.
In Sec. \ref{sec:results}, we present results for natural SUSY models
and compare them to results from unnatural SUSY models, and explain why
{\it natural SUSY} is more likely to emerge from anthropic selection
within the string landscape than unnatural SUSY models.
The answer is that there exists a substantial hypercube of parameter values
leading to $m_{weak}^{PU}\alt 4 m_{weak}^{OU}$ for natural SUSY models while the hypercube shrinks to relatively tiny volume for finetuned models: basically,
finetuning of parameters implies that only a tiny sliver of parameter choices
are likely to be phenomenologically (anthropically) allowed.
Some discussion and conclusions are presented in Sec. \ref{sec:conclude}.

\section{SUSY model}
\label{sec:nuhm}

\begin{figure}[t]
  \centering
  {\includegraphics[width=0.9\textwidth]{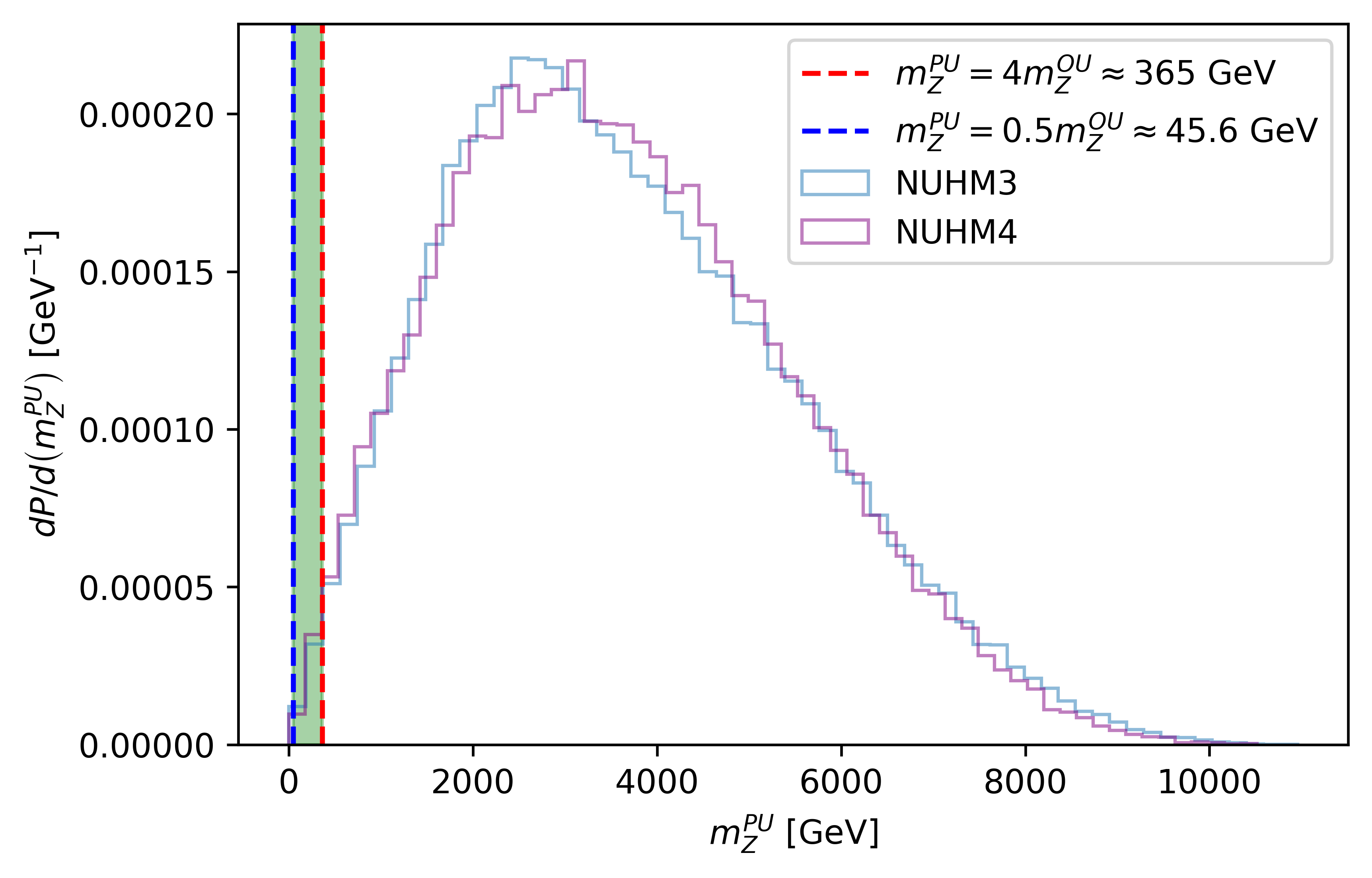}}\quad
  \caption{The predicted value of $m_Z^{PU}$ from scans over the NUHM3 and NUHM4 models if one doesn't fine-tune parameters to fix $m_Z^{PU}=m_Z^{OU}$.
 The green shaded band is the ABDS window.}
\label{fig:mZPU}
\end{figure}

In our present discussion, we will adopt the 2-3-4 extra parameter
non-universal Higgs model (NUHM2,3,4) for explicit calculations.
In this model, the matter scalars of the first two generations
are assumed to live in the 16-dimensional spinor of $SO(10)$
as is expected in string models exhibiting {\it local}
grand unification, where different gauge groups are present at different
locales on the compactified manifold\cite{Buchmuller:2005sh}.
In this case, it is really expected that each generation acquires a different
soft breaking mass $m_0(1)$, $m_0(2)$ and $m_0(3)$.
But for simplicity of presentation, sometimes we will assume generational
degeneracy. At first glance, one might expect that the generational
non-degeneracy would lead to violation of flavor-changing-neutral-current
(FCNC) bounds. The FCNC bounds mainly apply to first-second generation
nonuniversality\cite{Gabbiani:1996hi}. However, the landscape itself allows a solution to the SUSY
flavor problem in that it statistically pulls all generations to large values
provided they do not contribute too much to $m_{weak}^{PU}$. This means the 3rd
generation is pulled to $\sim$ several TeV values whilst first and second
generation scalars are pulled to values in the $10-50$ TeV range.
The first and second generation scalar contributions to the weak scale
are suppressed by their small Yukawa couplings\cite{Baer:2012cf}, whilst their $D$-term
contributions largely cancel under intra-generational universality\cite{Baer:2013jla}.
Their main influence on the weak scale then comes from two-loop RGE
contributions which, when large, suppress third generation soft term running
leading to tachyonic stop soft terms and possible charge-or-color breaking
(CCB) vacua which we anthropically veto\cite{Arkani-Hamed:1997opn,Baer:2000xa}.
These latter bounds are flavor independent so that first/second generation
soft terms are pulled to common upper bounds leading to a
quasi-degeneracy/decoupling solution to both the
SUSY flavor and CP problems\cite{Baer:2019zfl}.
Meanwhile, Higgs multiplets which live in different GUT representations
are expected to have independent soft masses $m_{H_u}$ and $m_{H_d}$\footnote{
  In models of local grand unification, the matter multiplets can live in the
  $SO(10)$ spinor reps while the Higgs and gauge fields live in split multiplets due to their geography on the compactified manifold\cite{Nilles:2014owa}.}.
Thus, we expect a parameter space of the NUHM models as
\be
m_0(1)\sim m_0(2),\ m_0(3),\ m_{H_u},\ m_{H_d},\ m_{1/2},\ A_0,\ B\mu,\ {\rm and}\ \mu . 
\ee
We assume that only models with appropriate electroweak symmetry breaking
(EWSB) are anthropically allowed (thus generating the weak scale), and the
scalar potential minimization conditions allow us to trade $B\mu$ for
$\tan\beta$. It is common practise to then finetune either $\mu$ or $m_{H_u}$
so as to generate $m_Z^{OU}=91.2$ GeV. But it is important that here we
{\it do not}
invoke this condition since we wish to allow $m_Z^{PU}$ to float to whatever
its derived value takes in the multiverse simulation.
For instance, the {\it predicted} value of $m_Z^{PU}$ from scans over the
NUHM3 and NUHM4 models is shown in Fig. \ref{fig:mZPU}. We also show the
ABDS window (shaded green in the Figure). The vast majority of models would be excluded since they lie
beyond the ABDS window (much like the vast majority of $\Lambda_{cc}$
in Weinberg's explanation of the cosmological constant). 

\section{Distribution of soft terms and $\mu$ parameter on the landscape}
\label{sec:soft}

\subsection{Soft terms}

How are the SUSY breaking soft terms expected to be distributed on the
landscape?
This information would be included in the landscape probability function
$f_{SUSY}(m_{soft})$.
A variety of proposals have been presented.
In Ref's \cite{Douglas:2004qg,Susskind:2004uv,Arkani-Hamed:2005zuc}, a power-law draw
\be
f_{SUSY}\sim m_{soft}^{2n_F+n_D-1}
\label{eq:fSUSY}
\ee
is expected where $n_F$ is the number of $F$-breaking fields and
$n_D$ is the number of $D$-breaking fields contributing to the overall
SUSY breaking scale $m_{soft}\sim m_{SUSY}^2/m_P$ where under gravity-mediation we
also expect the gravitino mass $m_{3/2}\sim m_{SUSY}^2/m_P$. The above form for
$f_{SUSY}$ arises if the SUSY breaking $F_i$ terms are distributed independently
as complex numbers on the landscape whilst the $D$-breaking fields $D_j$ are
distributed as real random numbers. Subsequently, it was then realized  that
the sources of SUSY breaking should not be all independent
which might spoil the above simplistic expectation\cite{Denef:2004cf}.
However, even under the condition of single
$F$-term source of SUSY breaking, then there is still a linear draw to
large soft terms $f_{SUSY}\sim m_{soft}^1$.
Furthermore, in Ref. \cite{Broeckel:2020fdz},
under considerations of K\"ahler moduli stabilization, then a {\it linear}
distribution $f_{SUSY}\sim m_{soft}^1$ would emerge from
KKLT\cite{Kachru:2003aw}-type moduli-stabilization.
In contrast, under dynamical SUSY breaking via
{\it e.g.} gaugino condensation\cite{Ferrara:1982qs} or instanton effects\cite{Affleck:1983rr},
all SUSY breaking scales are expected to be equally probable leading to
a distribution $f_{SUSY}\sim m_{soft}^{-1}$\cite{Dine:2004is,Dine:2005iw,Dine:2004ct}.
This distribution is also expected to emerge from
LVS-type\cite{Balasubramanian:2005zx} moduli-stabilization\cite{Broeckel:2020fdz}. This distribution favors smaller soft
terms and leads to sparticle masses below LHC search limits and
$m_h\ll 125$ GeV\cite{Baer:2021uxe} and so we will not consider it further here.

A final consideration is whether all soft terms should have common
probability distributions on the landscape\cite{Baer:2020vad}. For instance, gaugino
masses arise from the SUGRA gauge kinetic function $f_{AB}$ which is typically
of the form $\sim k\cdot S\delta_{AB}$ in string models where $S$ is the dilaton superfield and $k$ is some constant.
Under Eq. \ref{eq:fSUSY}, then this would give a linear draw to large
$m_{1/2}$ while scalar masses which arise from the K\"ahler function
might have a stronger draw to large values. In addition, the other soft
terms such as $A_0$ have very different dependencies on SUSY breaking fields
and hence are expected to scan independently on the landscape\cite{Baer:2020vad}.
For the bulk of this work, we will generally assume a single source of
SUSY breaking so all soft terms scan linearly in $m_{soft}$ to large values.

\subsection{$\mu$ term}

Since in this work we do not finetune the $\mu$ parameter to gain
the measured value of $m_Z^{OU}$, we must also be concerned with the
expected distribution $f_\mu (\mu )$. In solutions to the SUSY $\mu$
problem\cite{Bae:2019dgg}, it is usually expected that the $\mu$ term arises from SUSY
breaking via K\"ahler potential terms such as Giudice-Masiero\cite{Giudice:1988yz} (GM)
operators (wherein $\mu$ is expected to scan as do the soft terms) or via
superpotential terms such as $W\ni \lambda_\mu\phi^{n+1} H_uH_d/m_P^n$ where
$n=0$ gives the next-to-minimal MSSM\cite{Ellwanger:2009dp} (NMSSM) and $n>0$ gives the
Kim-Nilles\cite{Kim:1983dt} (KN) solution. Since Eq. \ref{eq:mzsPU} strongly favors
$\mu\sim m_{weak}\ll m_{soft}$ (the Little Hierarchy), then we expect
the GM solution disfavored as well. Likewise, we expect the NMSSM solution
to be disfavored since there is no evidence for visible sector singlet fields
which can re-introduce the gauge hierarchy\cite{Bagger:1995ay} or lead to domain wall
issues\cite{Abel:1995wk}.
Thus, we expect the KN mechanism, which may also be mixed
with the SUSY solution to the strong CP problem, as the most likely avenue
towards generating a superpotential $\mu$ term.
In Ref. \cite{Baer:2021vrk}, the distribution of $\mu$ terms under SUSY breaking
in the landscape was derived for fixed $\lambda_\mu$ values, where the
$\mu$ term was generated from the $Z_{24}^R$ discrete-$R$-symmetry
model\cite{Baer:2018avn} which generates the SUSY $\mu$-term of order $m_{weak}$ while
also generating a gravity-safe accidental, approximate global
Peccei-Quinn (PQ) symmetry needed to solve the strong CP problem in a stringy
setting where no global symmetries are allowed.
Instead of fixing $\lambda_\mu$ as in that work, we expect $\lambda_\mu$
to scan as would the other Yukawa couplings in the superpotential.
In Ref. \cite{Donoghue:2005cf}, Donoghue {\it et al.} showed that the distribution of
fermion masses are distributed uniformly on a log scale as expected
from the landscape. Since in our case the $\mu$ term also arises as a
superpotential Yukawa coupling, we will expect it to be scale-invariant
and hence distributed as
\be
f_\mu (\mu )\sim 1/\mu^{PU}
\ee
on the landscape (thus favoring small values of $\mu^{PU}$).

\section{The ABDS window}
\label{sec:ABDS}

\begin{figure}[t]
  \centering
  {\includegraphics[width=0.9\textwidth]{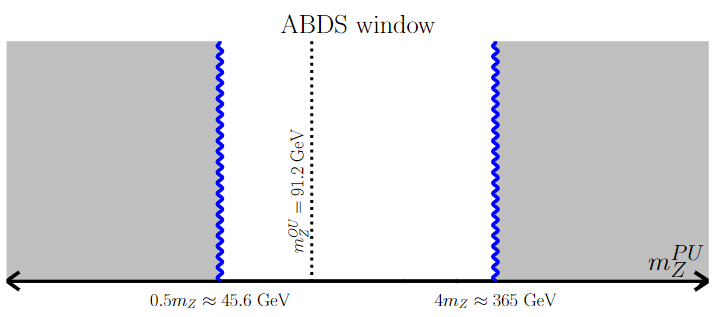}}\quad
  \caption{The ABDS-allowed window within the range of $m_Z^{PU}$ values.}  
\label{fig:ABDS}
\end{figure}

Agrawal {\it et al.}\cite{Agrawal:1997gf,Agrawal:1998xa} (ABDS) explored the plausible range of the weak
scale for pocket universes within the multiverse already in 1998.
They found that-- in order to allow the formation of
complex nuclei, and hence atoms as we know them which seem essential for life
to emerge-- the measured value of the weak
scale is located within a rather narrow window of values (the ABDS window).
Let us characterize the weak scale according to the oft-used $Z$-boson mass,
with $m_Z^{OU}=91.2$ GeV being the $Z$ mass in our universe while
$m_Z^{PU}\ne m_Z^{OU}$ is the $Z$-boson mass in each different pocket universe.
ABDS assumed an ensemble of pocket universes with the SM as the $4-d$ EFT
but with variable values of $m_Z^{PU}$. Then, ABDS found $m_Z^{PU}\alt (2-5)m_Z^{OU}$ as an upper bound, while (for us) the less essential lower bound is
$m_Z^{PU}\agt m_Z^{OU}/2$.
The ABDS pocket universe window is depicted in Fig. \ref{fig:ABDS}.

While the exact upper bound on $m_Z^{PU}$ is not certain, what is important
is that there is some {\it distinct upper bound}: for too large of values of
$m_Z^{PU}$, then complex stable nuclei will not form, and hence the
complex chemistry of our own universe will also not form. The explicit
ABDS upper bound is quite distinct from many previous approaches which would
penalize too large values of $m_Z^{PU}$ by a factor $(m_{weak}^{OU}/m_{SUSY})^2$,
a putative finetuning factor which penalizes but does not disallow a large
mass gap (a Little Hierarchy) between the SUSY breaking scale and the
measured value of the weak scale.

\begin{figure}[t]
  \centering
  {\includegraphics[width=1.0\textwidth]{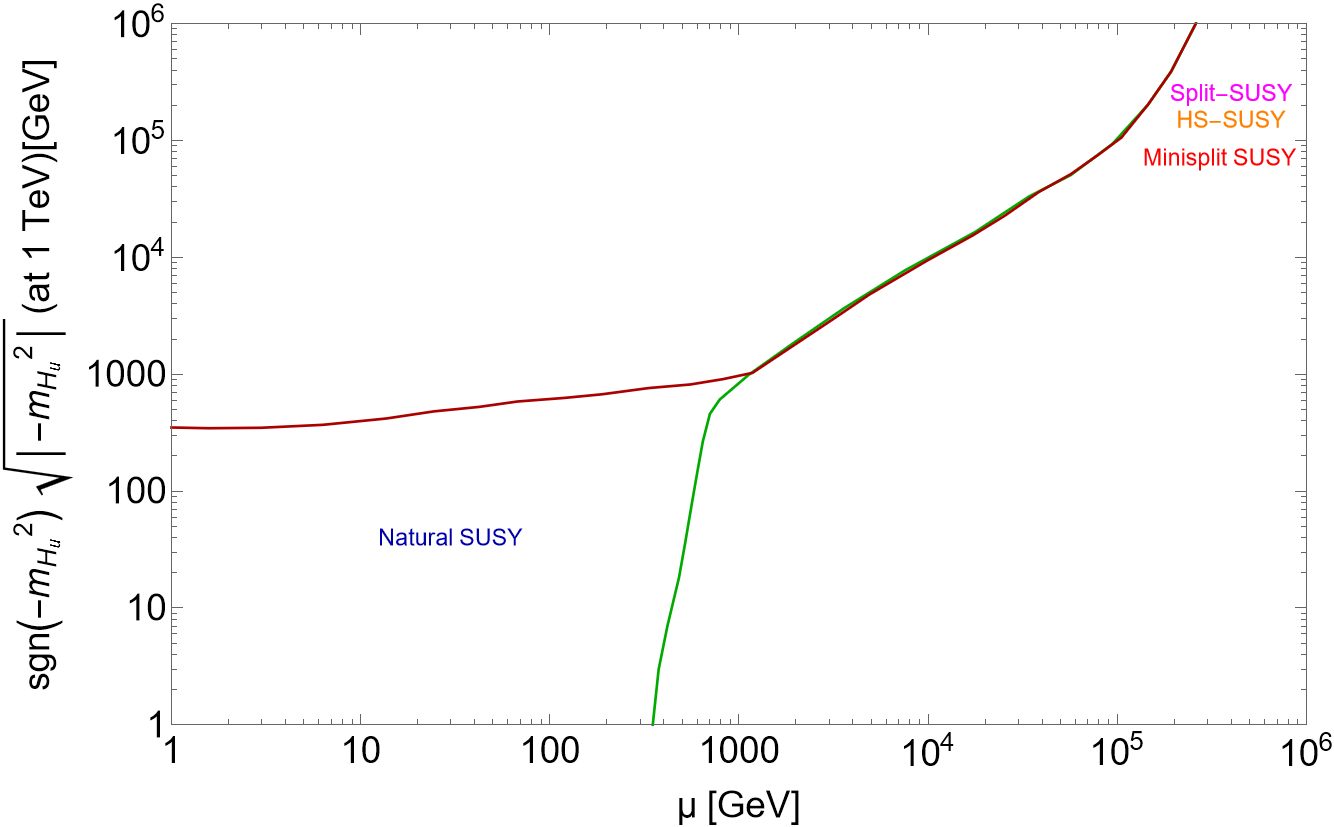}}\quad
  \caption{The $\mu^{PU}$ vs. $\sqrt{-m_{H_u}^2(weak)}$ parameter
space in a toy model ignoring radiative corrections to the
Higgs potential. The region between red and green curves
leads to $m_{weak}^{PU}<4 m_{weak}^{OU}$ so that the atomic principle 
is satisfied.
}  
\label{fig:mu_mhu}
\end{figure}

Using conventional SUSY spectra generators where $m_Z$ is fixed to its
measured value, then one may estimate the value of $m_Z^{PU}$ from the
finetuning measure $\Delta_{EW}$ in the limit of no finetuning (where
the weak scale is determined by the maximal value on the RHS of
Eq. \ref{eq:mzsPU}) as
\be
m_Z^{PU}\simeq \sqrt{\Delta_{EW}/2}m_Z^{OU}
\ee
which gives for $\Delta_{EW}\sim 30$ a value $m_Z^{PU}\sim 360$ GeV,
about four times its value $m_Z^{OU}$.
We can then plot out the allowed range of MSSM weak scale parameters
from Eq. \ref{eq:mzsPU} while setting the radiative corrections
$\Sigma_u^u$ and $\Sigma_d^d$ to zero. The result is shown in Fig. \ref{fig:mu_mhu} where we plot allowed values of $\mu^{PU}$ vs. $\sqrt{-m_{H_u}^2(weak)}$.
To fulfill  the requirement that $m_Z^{PU}\alt 4m_Z^{OU}$, then one must live
in between the red and green curves. For parameter choices above the red curve
then one obtains $m_Z^{PU}\gg m_Z^{OU}$ and the pocket universe value of the
weak scale is too big. For points below the green curve, then $m_Z^{PU2}$
goes negative signalling inappropriate EWSB.

One immediately notices that there is a large range of parameter values
in the lower-left corner of the plot that land in the ABDS window.
Alternatively, for large values of $\mu^{PU}\agt 360$ GeV and
$\sqrt{-m_{H_u}^2(weak)}\agt 360$ GeV, then in order to gain $m_Z^{PU}$
in the ABDS window one must land within the tiny gap between the
red and green curves. Unnatural or finetuned SUSY models
(such as High-Scale SUSY\cite{Elor:2009jp,Giudice:2011cg}, Split SUSY\cite{Arkani-Hamed:2004ymt,Arkani-Hamed:2004zhs} and Minisplit SUSY\cite{Arvanitaki:2012ps,Arkani-Hamed:2012fhg}, labelled in upper right) thus must fall in
the narrow gap whilst natural SUSY models characterized by low $\Delta_{EW}$
would lie in the substantial lower-left allowed region.
If $\mu^{PU}$ and $\sqrt{-m_{H_u}^2(weak)}$ were fundamental parameters (as in pMSSM) that were distributed in a scale invariant fashion (uniform on a log scale), then it would be easy to see why natural SUSY is more likely to emerge
than finetuned SUSY from the landscape: for a random distribution of parameters
(on a log scale), one is more likely to land in the large lower-left region than in the narrow gap between the red and green curves in the upper right. 

Of course, the value of $m_{H_u}^2(weak)$ is highly distorted from its high scale
value due to RG running and the requirement of radiatively-induced EWSB (REWSB).
However, the value of $\mu^{PU}$ roughly tracks the high scale value of
$\mu$ since $\mu$ does not run that much between high and low scales.
In addition, the $\Sigma_{u,d}^{u,d}$ terms are not zero and can often be the
dominant contributions to the RHS of Eq. \ref{eq:mzsPU}.
Our goal in this paper is to explore these connections numerically
via a toy simulation of the multiverse.

\subsection{The hypercube of ABDS-allowed parameter values in the NUHM2 model}

Before we turn to our toy multiverse simulation, let us illustrate
the hypercube of ABDS-allowed parameters for the simple case of the NUHM2
model. While a multi-dimensional portrayal of the hypercube
is not possible, here we show $1-d$
parameter portrayals for the case where all other parameters are fixed.
Thus, we adopt a
NUHM2 benchmark model with $m_0=5$ TeV, $m_{1/2}=1.2$ TeV, $A_0=-8$ TeV,
$\mu =200$ GeV and $m_A=2$ TeV with $\tan\beta =10$.
This model has $m_h=124.7$ GeV with $\Delta_{EW}=22$ from the Isasugra
spectrum generator\cite{Paige:2003mg}. 
\begin{figure}[H]
\begin{center}
\includegraphics[height=0.22\textheight]{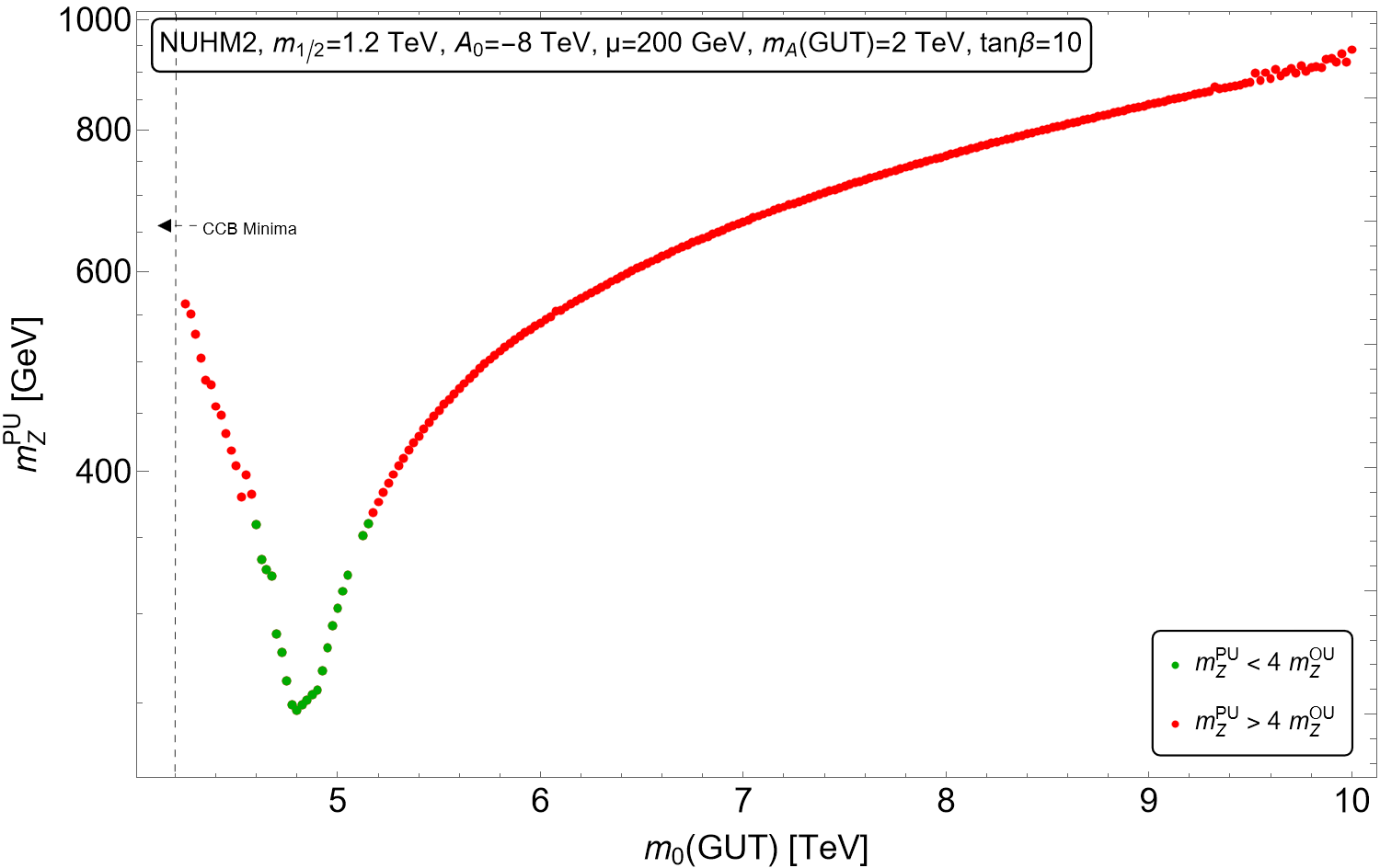}
\includegraphics[height=0.22\textheight]{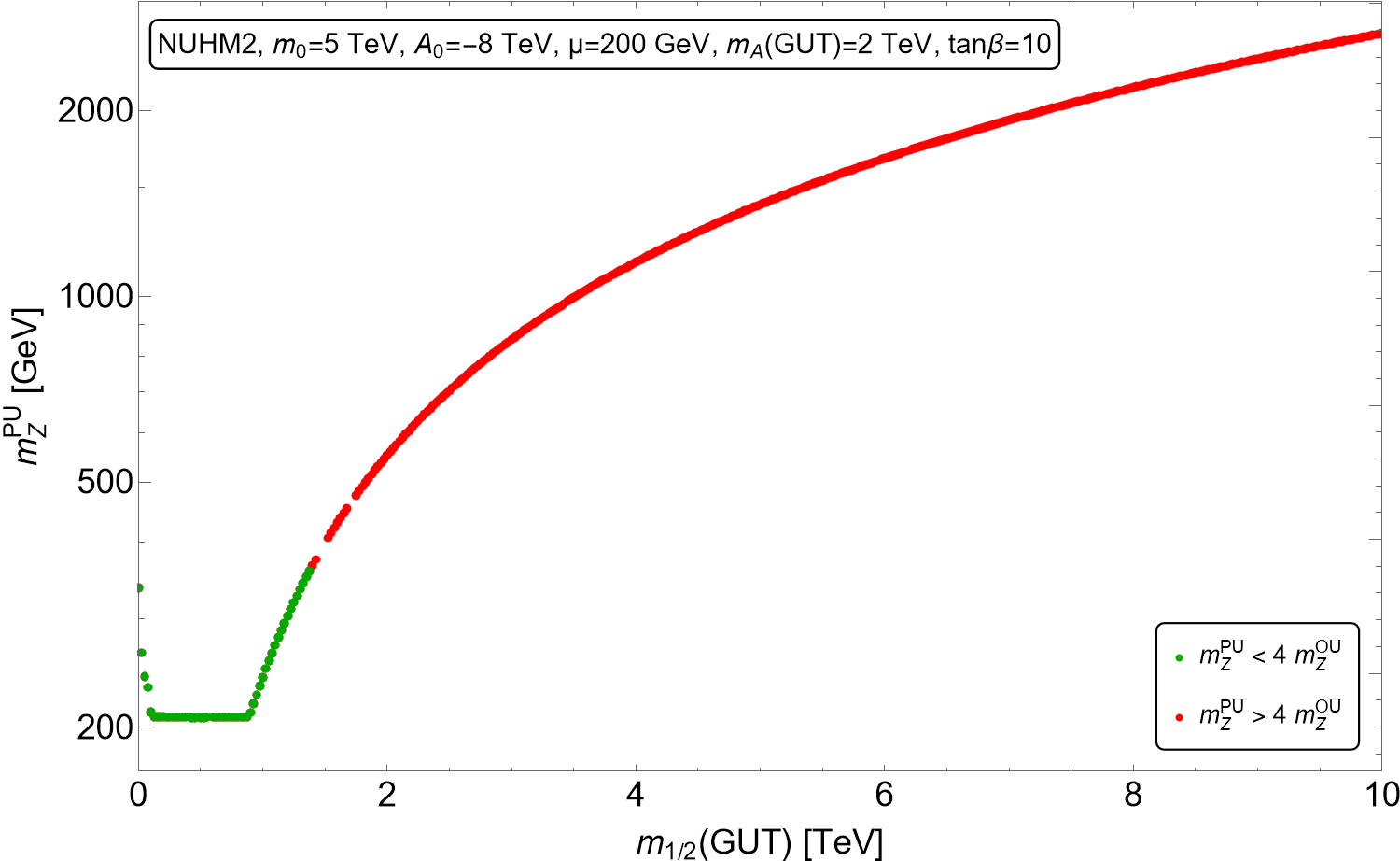}\\
\includegraphics[height=0.22\textheight]{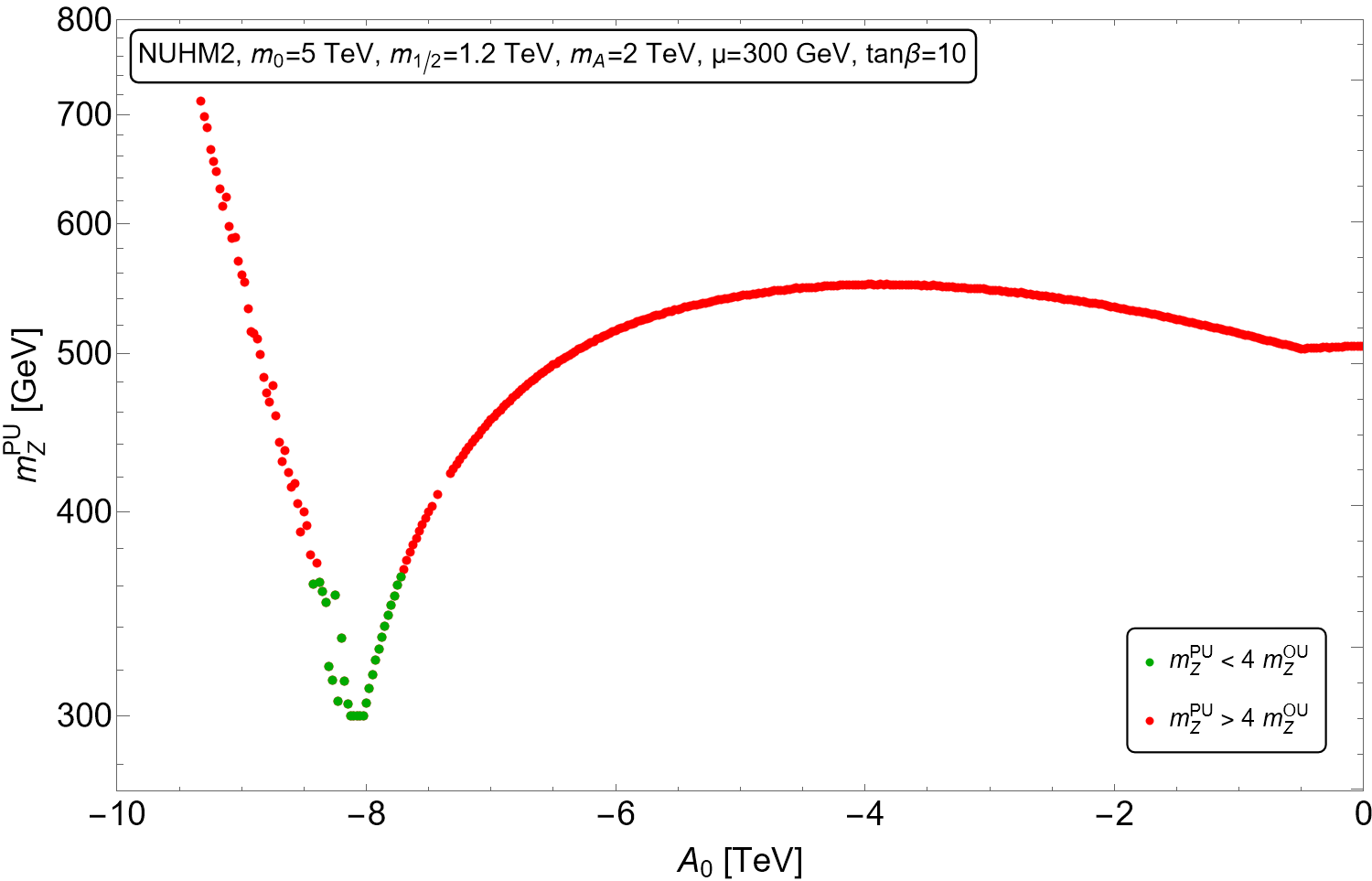}
\includegraphics[height=0.22\textheight]{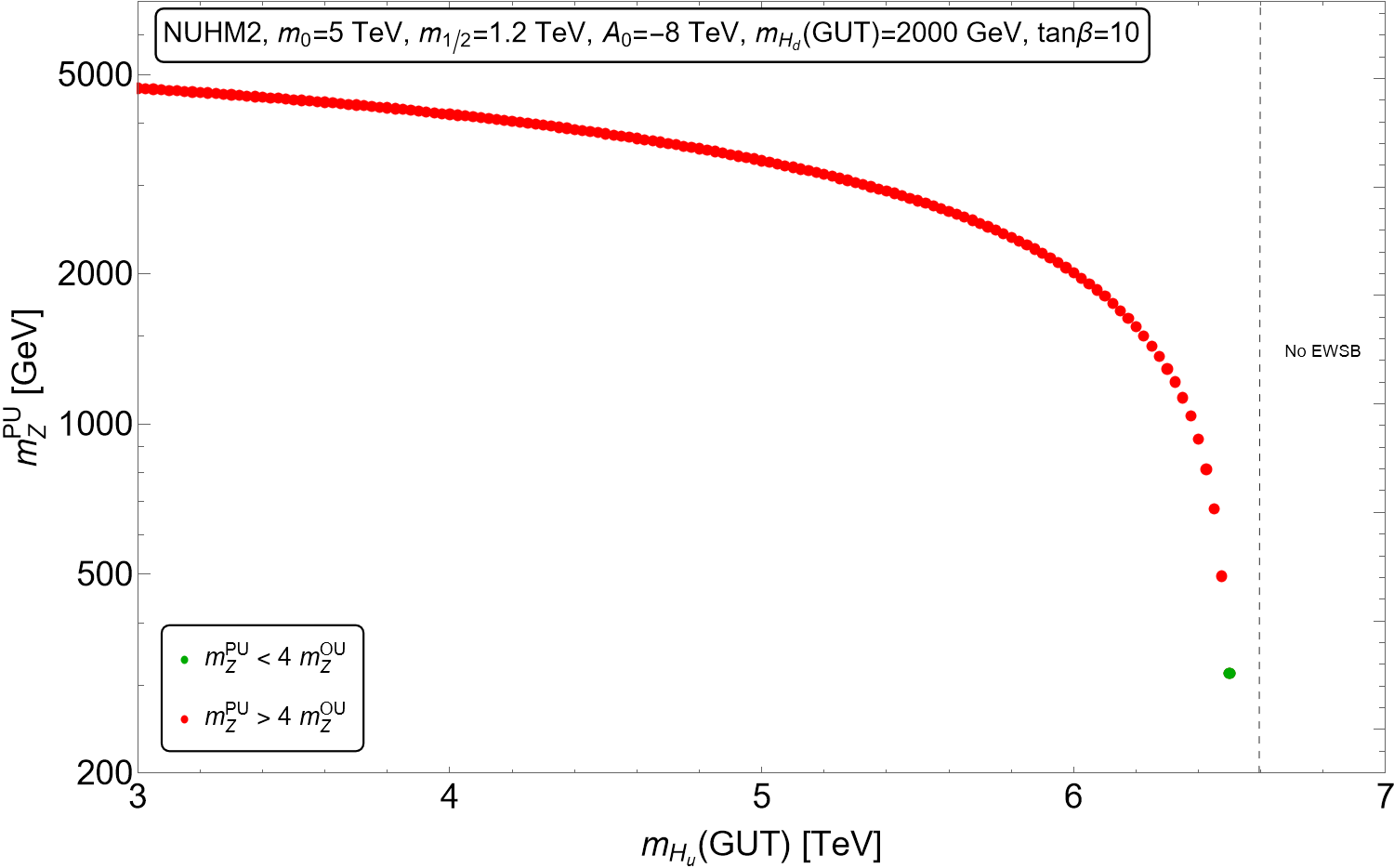}
\caption{The value of $m_{weak}^{PU}$ vs. various NUHM2 model
  parameters to illustrate the substantial hypercube of parameter
  values which lead to $m_Z^{PU}$ within the ABDS window.
The green points denote vacua with appropriate
EWSB and with $m_{weak}^{PU}<4 m_{weak}^{OU}$ so that the atomic principle 
is satisfied. Red points have $m_{weak}^{PU}>4 m_{weak}^{OU}$.
Parameter values include
{\it a}) $m_0(1,2,3)$, {\it b}) $m_{1/2}$, {\it c}) $A_0$ and
{\it d}) $m_{H_u}$.
\label{fig:hypercube}}
\end{center}
\end{figure}

In Fig. \ref{fig:hypercube}, we plot the value of $m_Z^{PU}$ versus
variation in several soft SUSY breaking terms for a NUHM2 benchmark model.
We take $m_Z^{PU}=\sqrt{\Delta_{EW}/2}m_Z^{OU}$.
In frame {\it a}), we vary the parameter $m_0$.
The red dots correspond to $m_Z^{PU}>4 m_Z^{OU}$ while green points have
$m_Z^{PU}<4m_Z^{OU}$. From the plot, we see range of $m_0:4.5-5.2$ TeV
which leads to ABDS-allowed pocket universes.
For larger or smaller values, then the $\Sigma_u^u(\tst_{1,2})$ values
which enter Eq. \ref{eq:mzsPU} become too large and then finetuning is required to lie within the ABDS window.
This range of allowed $m_0$ values is correlated with the window of
$A_0$ values in that large cancellations can occur in both $\Sigma_u^u(\tst_1 )$
and $\Sigma_u^u(\tst_2 )$ for large $A_0$ at nearly maximal stop mixing
which is where $m_h$ is lifted to $\sim 125$ GeV\cite{Baer:2012up}.
For lower or higher $m_0$ values, this cancellation is destroyed and
top-squark loop contributions to the weak scale become too large.

In frame {\it b}), we show $m_Z^{PU}$ vs. variation in
unified gaugino mass $m_{1/2}$. For very low $m_{1/2}$, then the $\mu$ term
gives the dominant contribution to $m_Z^{PU}$. But as $m_{1/2}$ increases, then
the top-squark contributions $\Sigma_u^u(\tst_{1,2})$ become large.
Requiring $m_Z^{PU}\alt 4 m_Z^{OU}$ places an upper bound on $m_{1/2}$,
in this case around $1.5$ TeV. Thus, the window in $m_{1/2}$ lies
between $0-1.5$ TeV.

In frame {\it c}), we show variation in $m_Z^{PU}$ vs. $-A_0$
(the negative values give cancellations in $\Sigma_u^u$ around the same
values of $A_0$ which lift $m_h\sim 125$ GeV). Here we see the
ABDS-allowed window extends from $\sim -7.5$ TeV to $\sim -8.5$ TeV;
for this range, the $\Sigma_u^u(\tst_{1,2})$ contributions to $m_Z^{PU}$
are suppressed by cancellations.

\begin{figure}[t]
  \centering
  {\includegraphics[width=0.8\textwidth]{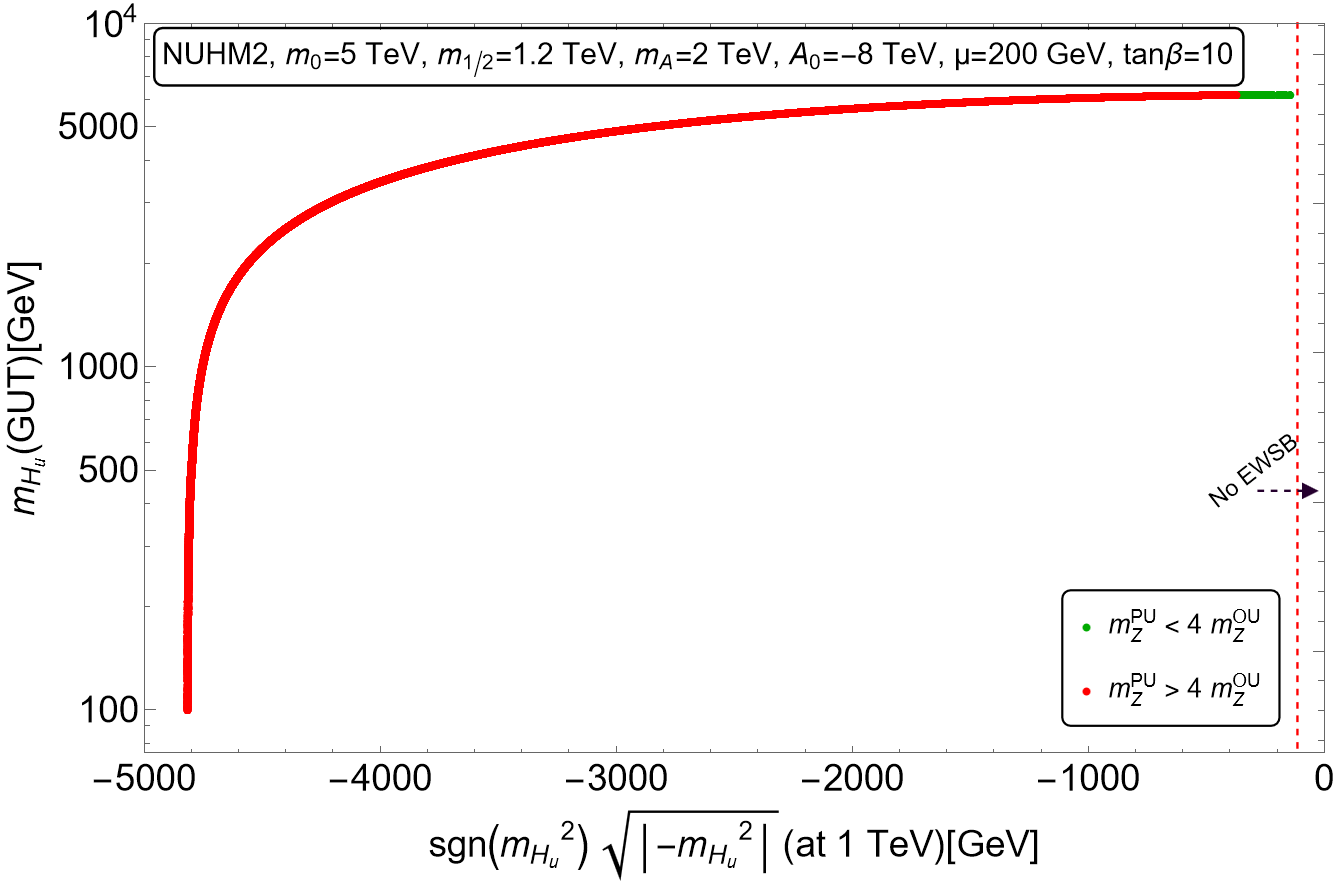}}\quad
  \caption{The value of $m_{H_u}(weak)\equiv sign(m_{H_u}^2)\sqrt{-m_{H_u}^2(weak)}$ vs. $m_{H_u}(m_{GUT})$
The green points denote vacua with appropriate
EWSB and with $m_{weak}^{PU}<4 m_{weak}^{OU}$ so that the atomic principle 
is satisfied. Red points have $m_{weak}^{PU}>4 m_{weak}^{OU}$.
}  
\label{fig:mHu_mHuGUT}
\end{figure}

Finally, in frame {\it d}), we show variation of $m_Z^{PU}$ vs.
variation in $m_{H_u}$. In this case, for too small of values of
$m_{H_u}(GUT)$, then the value of $m_{H_u}^2(weak)$ is driven to large negative
values (see {\it e.g.} Fig. 4 of Ref. \cite{Baer:2019cae}), and hence
gives a large contribution to $m_Z^{PU}$ via Eq.~\ref{eq:mzsPU}.
As $m_{H_u}(GUT)$ increases, then $-m_{H_u}^2(weak)$ decreases, until
at $m_{H_u}(GUT)\sim 6.5$ TeV, then electroweak symmetry is no longer
broken and we do not generate a weak scale.
Such pocket universes must be vetoed since they lack massive SM fermions
and gauge bosons. The landscape pull on $m_{H_u}$ is to large values such that
electroweak symmetry (EWS) is barely broken.
While this viable portion of the hypercube looks small, it must be
remembered that there is a landscape pull to large values stopping just short
of the no EWSB limit (dashed vertical line). 
We can view this differently in Fig. \ref{fig:mHu_mHuGUT} where instead
we plot $m_{H_u}(GUT)$ on the $y$-axis and $m_{H_u}(weak)$ on the $x$-axis.
In this case, the more substantial allowed range of
$m_{H_u}(weak)$ values required by Fig. \ref{fig:mu_mhu} is apparent
as the green region on the right side of the curve.

\section{A toy model of vacuum selection within the multiverse}
\label{sec:model}

Instead of using any of the publicly available SUSY spectra codes for
which $m_Z$ is fixed at its measured value in our universe, we will
construct a toy program with variable weak scale where both $\mu$ and
$m_{H_u}$ are input parameters and $m_Z^{PU}\ne m_Z^{OU}$ is an output parameter.
We begin by creating a code which solves the 26 coupled renormalization
group equations (RGEs) of the MSSM via Runge-Kutta method starting with
GUT scale inputs of parameters
\be
m_0(1,2,3),\ m_{H_u},\ m_{H_d},\ m_{1/2},\ A_0,\ \tan\beta ,\ {\rm and}\ \mu
\ee
where we have used the EWSB minimization conditions to trade
the bilinear soft term $b=B\mu$ for $\tan\beta$, but where we have not
imposed the relation between $m_{H_u}(weak)$ and $\mu (weak)$ in terms
of the measured value of $m_Z$. We use the one-loop RGEs but augmented
by the two-loop terms from Eq. 11.22 of Ref. \cite{Baer:2006rs} which set the upper limits
on first/second generation scalar masses. We run the set of soft terms,
gauge and Yukawa couplings and $\mu$ term from $Q=m_{GUT}\simeq 2\times 10^{16}$ GeV
down to the weak scale $Q_{weak}$ which we define as that scale at which
$m_{H_u}^2$ first runs negative so long as $Q_{weak}<m_0(3)$. Otherwise, we
set $Q_{weak}=m_0(3)$. This method implements the condition of
{\it barely broken EWSB}\cite{Giudice:2006sn}. Then, we use Eq. \ref{eq:mzsPU} to
calculate $m_Z^{PU}$ to see if it lies within the ABDS window.
We veto vacua with no EWSB or color-or-charge-breaking (CCB) minima
(where charged or colored scalar squared masses run negative) as these
would presumably lead to unlivable vacua.

In our toy simulation of this  fertile patch (those vacua leading to the
MSSM as the $4-d$ low energy EFT) of the string landscape, we will
scan over parameters as such:
\bea
m_0(1,2)&:& 0-60\ {\rm TeV}\\
m_0(3)&:& 0.1-10\ {\rm TeV}\\
m_{H_u}&:& m_0(3)-2m_0(3)\\
m_{H_d}(\sim m_A) &:& 0.3-10\ {\rm TeV}\\
m_{1/2}&:& 0.5-3\ {\rm TeV}\\
-A_0&:& 0-50\ {\rm TeV}\\
\mu_{GUT}&:&1-10^4\ {\rm TeV}\\
\tan\beta &:& 3-60
\eea
The soft terms are all scanned according to  $f_{SUSY}\sim m_{soft}^1$
(as expected for SUSY breaking from a single $F$-term field)
while $\mu$ is scanned according to $f_\mu\sim 1/\mu$.
For $\tan\beta$, we scan uniformly.

\section{Numerical results}
\label{sec:results}

\subsection{Results for $m_{H_u}$ vs. $\mu$ plane}

We are now ready to present results from our toy model simulation of
vacuum selection from the multiverse, where for simplicity we restrict
ourselves to those vacua with the MSSM as the low energy EFT, but where
soft terms and the $\mu$ parameter vary from vacuum to vacuum, and with a
linear draw to large soft terms (as expected in models with
spontaneous supersymmetry breaking from a single $F$-term field where
{\it all} field values are equally likely). In this case, since $\langle F\rangle$ is
distributed randomly as a complex number, then the overall SUSY breaking
scale $m_{SUSY}$ has a linear draw to large soft terms\cite{Baer:2016lpj}.
We couple this
with the MSSM prediction for the magnitude of the weak scale as given by
Eq. \ref{eq:mzsPU}. This is one of the most important predictions of
supersymmetric models. However, it is often hidden in phenomenological
work since parameters are tuned in the computer codes so that the
value of $m_Z$ has its numerical value as given in our universe.

\begin{figure}[t]
  \centering
  {\includegraphics[width=1.0\textwidth]{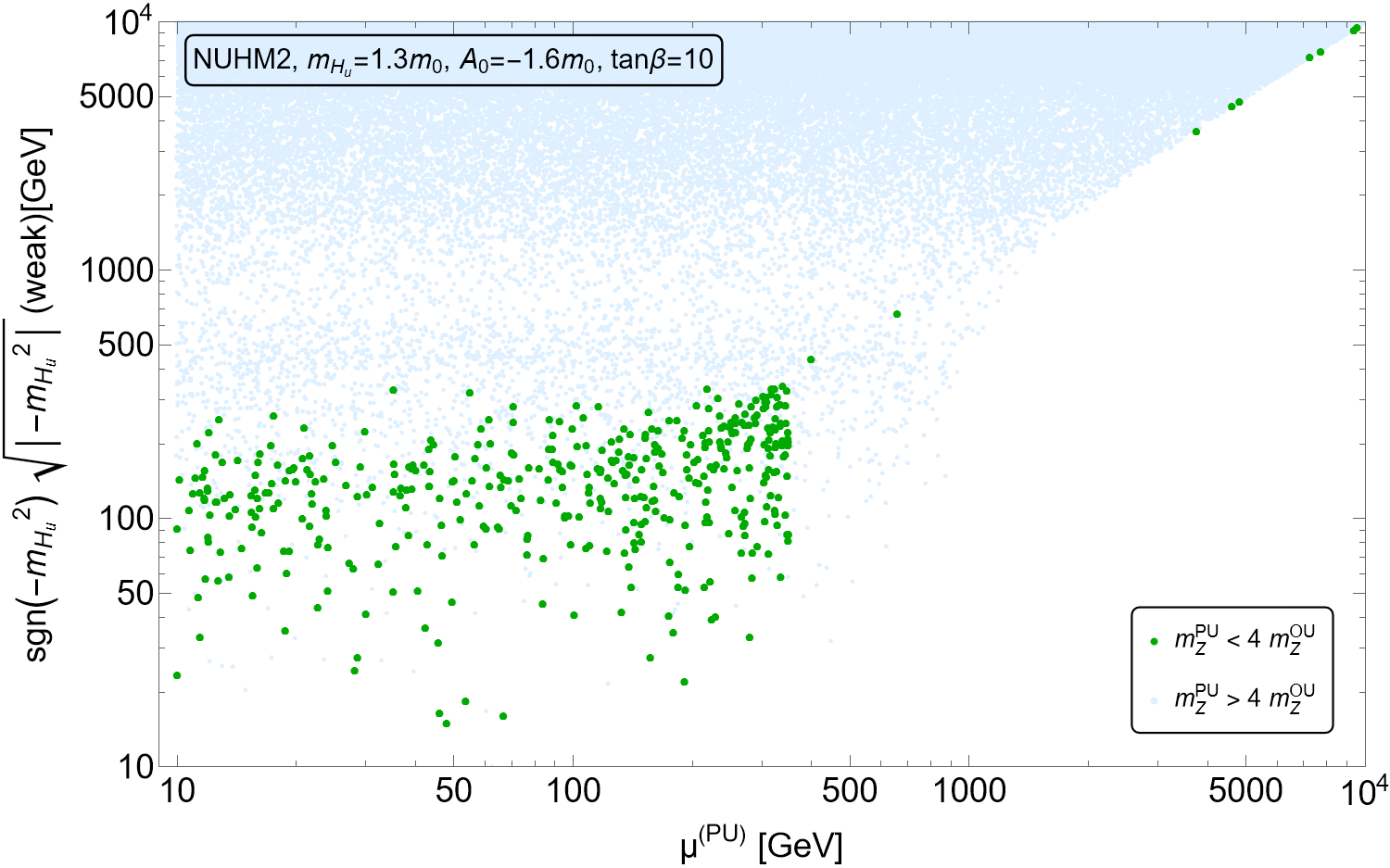}}\quad
  \caption{The value of $m_{H_u}(weak)$ vs. $\mu^{PU}$
The green points denote vacua with appropriate
EWSB and with $m_{weak}^{PU}<4 m_{weak}^{OU}$ so that the atomic principle 
is satisfied. Blue points have $m_{weak}^{PU}>4 m_{weak}^{OU}$.
}  
\label{fig:dots1}
\end{figure}

In Fig. \ref{fig:dots1}, we show the results of our toy model where
$m_Z^{PU}\ne m_Z^{OU}$. We show results in the $\mu^{PU}$ vs.
$sign(-m_{H_u}^2)\sqrt{|-m_{H_u}^2|}(weak)$ parameter plane as in
Fig. \ref{fig:mu_mhu}.
We adopt parameter choices $m_{H_u}=1.3 m_0$ and $A_0=-1.6 m_0$ with $\tan\beta =10$ while allowing $m_0$ and $m_{1/2}$ to be statistically determined
as in Sec. \ref{sec:model}.
The soft term $m_{H_d}$ is scanned over the range given above with $n=1$.
The light blue points all have
$m_Z^{PU}>4 m_Z^{OU}$ and so lie beyond the ABDS window: the weak scale is too
large to allow for formation of complex nuclei and hence atoms as we know them:
these points are anthropically vetoed. The green points have values of
$m_Z^{PU}<4 m_Z^{OU}$ and hence fall within the ABDS window: these points should
allow for formation of complex nuclei and obey the atomic principle\cite{Arkani-Hamed:2005zuc}. We see that the bulk of allowed points live within the
parameter hypercube as shown in Fig. \ref{fig:mu_mhu}.
However, in our toy model, a small number of green points now do live beyond the
Fig. \ref{fig:mu_mhu} parameter hypercube. The latter points are generated with
accidental finetuning of parameters such that $m_Z^{PU}$ is still
less than $4 m_Z^{OU}$ in spite of large contributions to the weak scale.
Nonetheless, we do see that the natural SUSY models with $\mu^{PU}$
and $m_{H_u}(weak)\alt 360$ GeV are much more numerous than the
finetuned solutions.

\begin{figure}[t]
  \centering
  {\includegraphics[width=1.0\textwidth]{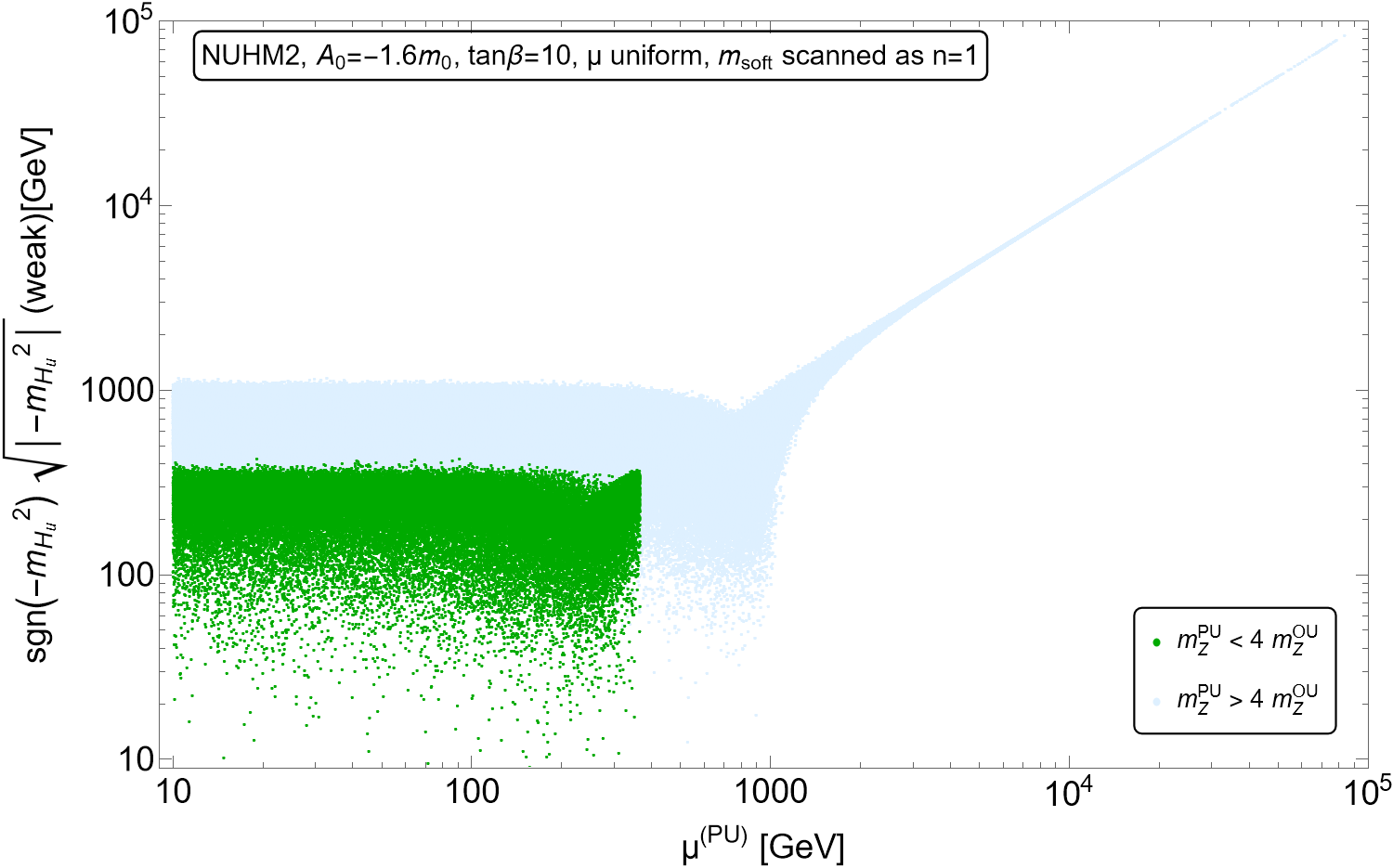}}\quad
  \caption{Points with $m_Z^{PU}<4m_Z^{OU}$ (green) from Isajet
with $m_Z^{PU}<4 m_Z^{OU}$ (green) and with $m_Z^{PU}>4m_Z^{OU}$ (blue) 
from the NUHM2 model. This run implements $m_Z^{PU}=\sqrt{\Delta_{EW}/2} m_Z^{OU}$.
}  
\label{fig:dew}
\end{figure}

Let us compare the above results to the more common methodology of
simply requiring $m_Z^{PU}=\sqrt{\Delta_{EW}/2}m_Z^{OU}<4m_Z^{OU}\simeq 360$ GeV
(corresponding to $\Delta_{EW}<30$) as can be
computed in available SUSY spectrum generators\footnote{The new code
  DEW4SLHA\cite{Baer:2021tta} allows one to compute $\Delta_{EW}$ from the
  SUSY Les Houches Accord (SLHA) output files of any of the available SUSY
  spectrum generators.}.
We show these results as a scan in the same parameter space as in
Fig. \ref{fig:mu_mhu} but now as shown in Fig. \ref{fig:dew}.
In this case, the weak scale is taken to be the largest of the elements
contributing to the RHS of Eq. \ref{eq:mzsPU}, so it assumes no finetuning
of parameters within the landscape vacuum states. Again, the blue points lie
beyond the ABDS window whilst the green points are anthropically allowed.
In this case, the green points fill out the parameter hypercube of
Fig. \ref{fig:mu_mhu}, albeit including the $\Sigma_{u,d}^{u,d}$
radiative corrections. Since no allowance for finetuning is made, then
no green points extend along the finetuned diagonal in Fig. \ref{fig:dew}.
The plot does show why the event generator runs with
$f_{EWSB}=\Theta (30-\Delta_{EW})$ gives a good representation of
expected superparticle and Higgs mass spectra in scans over the landscape
of string vacua\cite{Baer:2017uvn,Baer:2019tee,Baer:2020dri}.

\subsection{Distribution of $\mu$ parameter}

As a byproduct of our toy model of the string landscape, we are able to plot out
the expected distribution of the superpotential $\mu$ parameter. This has also
been done in Ref. \cite{Baer:2021vrk} but in that case a fixed $\mu$-term Yukawa
coupling $\lambda_\mu$ is adopted for a particularly well-motivated
solution to the $\mu$ problem wherein the global $U(1)_{PQ}$ symmetry
needed to solve the strong CP problem emerges as an
accidental, approximate, gravity-safe  global symmetry from a
discrete anomaly-free $R$-symmetry ${\bf Z}_{24}^R$ which also
solves the SUSY $\mu$ problem and provides a basis for $R$-parity\cite{Baer:2018avn}.
In the present case, we allow $\lambda_\mu$ to also scan in the
landscape so that $\mu$ is distributed uniformly across the decades
of possible values, as may be expected for other superpotential terms
(the matter Yukawa couplings) as shown by Donoghue {\it et al.}\cite{Donoghue:2005cf}.

\begin{figure}[t]
  \centering
        {\includegraphics[width=1.0\textwidth]{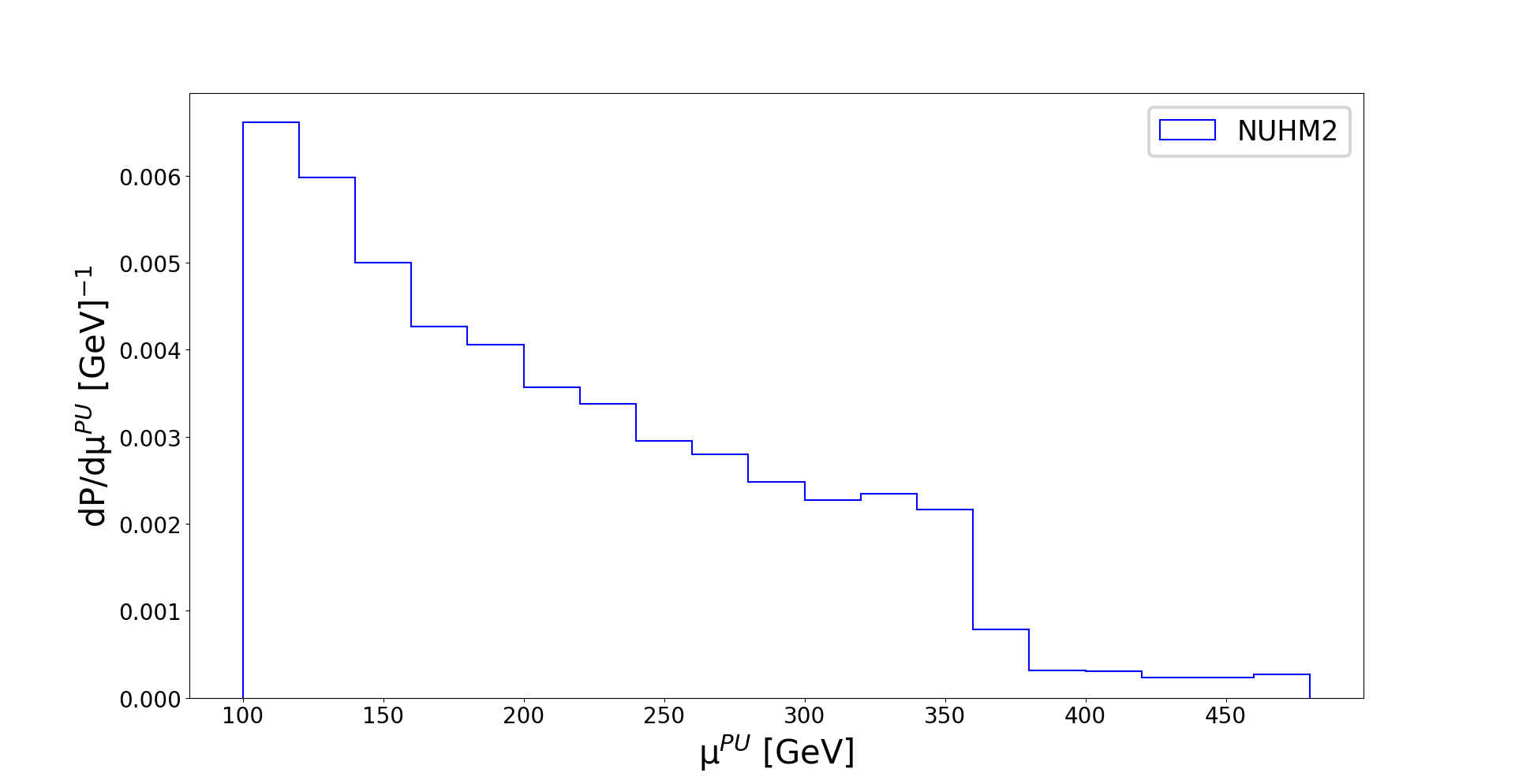}}\quad
  \caption{Distribution of $\mu$ parameter from scan over NUHM2
    model with $P_\mu\sim 1/\mu$ and with $m_Z^{PU}<4 m_Z^{OU}$
    and with other parameters fixed as in Fig. \ref{fig:dots1}.}  
\label{fig:mu}
\end{figure}

In Fig. \ref{fig:mu}, we show the distribution of the weak scale value of
the SUSY $\mu$ parameter as expected from our toy landscape model where
$\mu^{PU}$ is distributed as $P_\mu\sim 1/\mu$ and where we also
require appropriate EWSB and $m_Z^{PU}<4 m_Z^{OU}$.
Other parameters are fixed as in Fig. \ref{fig:dots1}.
From the plot, we see
that the $\mu^{PU}$ distribution is peaked at low values and falls off
at higher $\mu^{PU}$ values of several hundred GeV, with a steep
drop beyond the non-finetuned ABDS window which ends at $\mu\sim 360$ GeV.
For our toy model, there
is still some probability to gain $\mu^{PU}\agt 360$ GeV due to the
possibility of finetuning in our toy model. It should be noted that the lower
range of $\mu$ values $\sim 100-200$ GeV is now under pressure from
LHC soft dilepton plus jets plus MET searches\cite{ATLAS:2019lng,CMS:2021xji}
via higgsino pair production\cite{Han:2014kaa,Baer:2014kya,Baer:2020sgm,Baer:2021srt}.
The plot also emphasizes that there would be a good chance that ILC would
turn out to be a {\it higgsino factory} in addition to a Higgs factory for
$\sqrt{s}\agt 2m(higgsino)$\cite{Baer:2014yta,Baer:2019gvu}.
%

%
%

%
%

\subsection{Why finetuned SUSY models are scarce on the landscape compared to natural SUSY}

In Fig. \ref{fig:dots1}, we see that finetuned SUSY models that lie
with parameters $\mu^{PU}$ and $m_{H_u}(weak)\gg 4m_Z^{OU}$ are
relatively scarce in the multiverse compared to natural SUSY models with
low $\Delta_{EW}$. While the finetuned models are logically possible,
selection of their parameters is restricted to a hypercube of tiny-volume
compared to natural SUSY models, and so we expect natural SUSY
as the more likely expression of anthropically selected pocket universes.

\begin{figure}[tbh!]
  \centering
  {\includegraphics[width=0.6\textwidth]{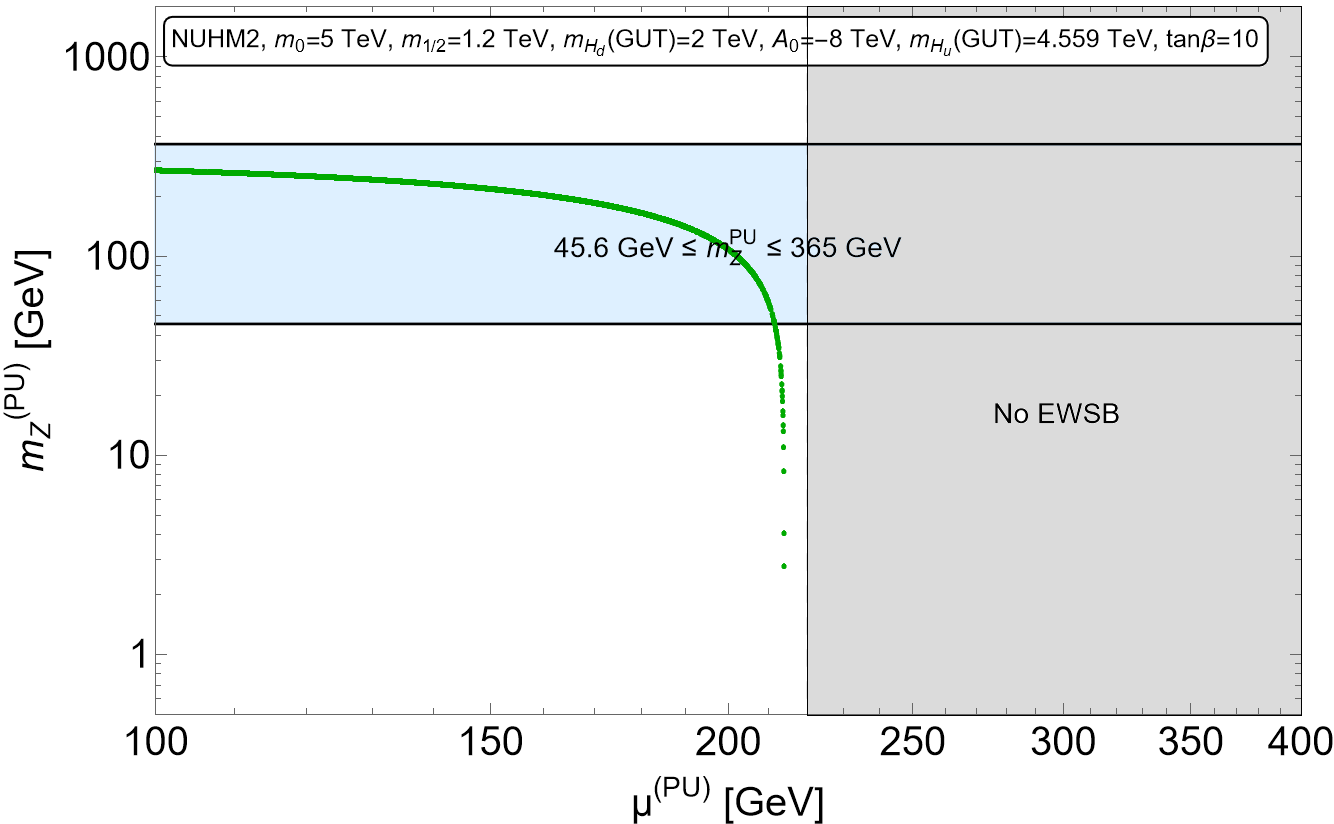}}\\
  {\includegraphics[width=0.6\textwidth]{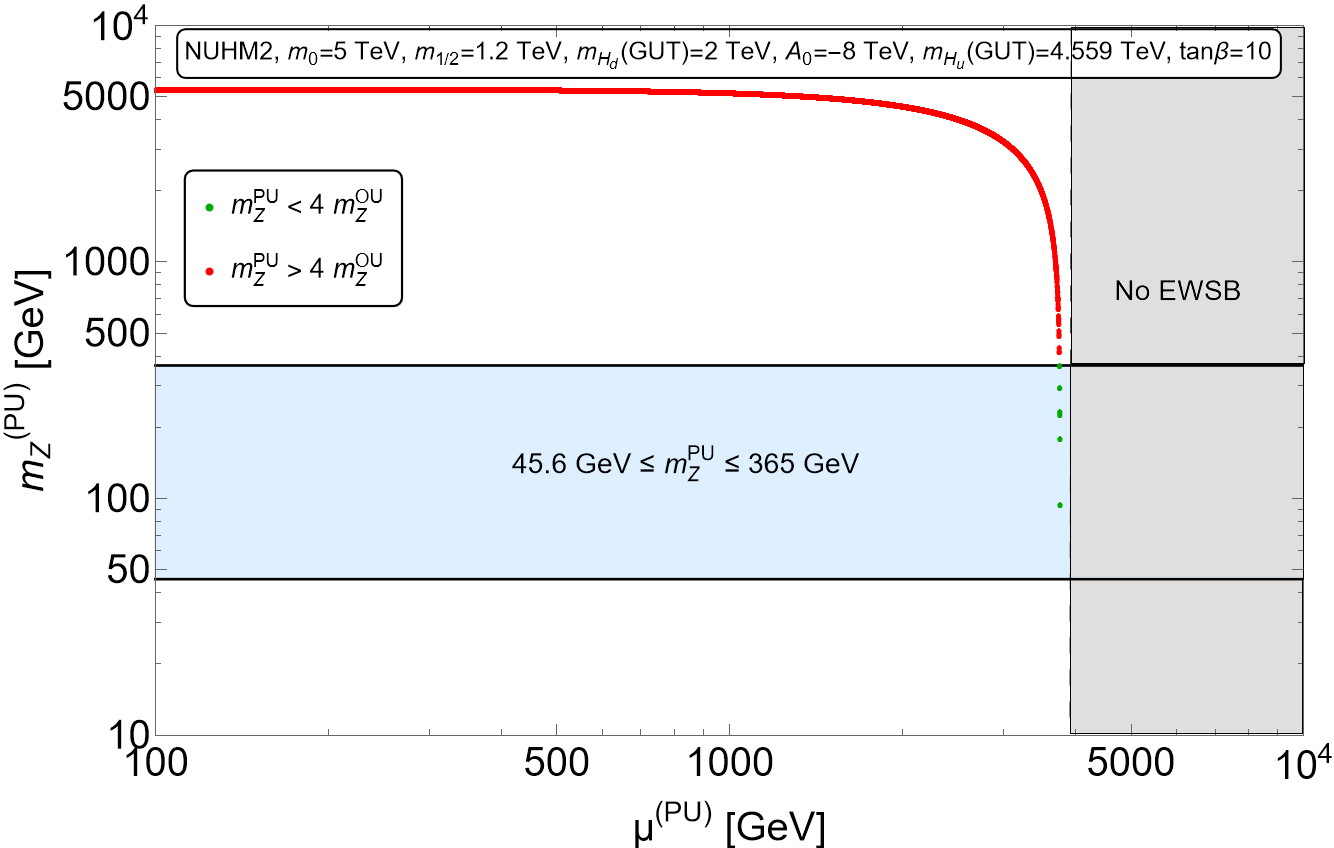}}\\
    {\includegraphics[width=0.6\textwidth]{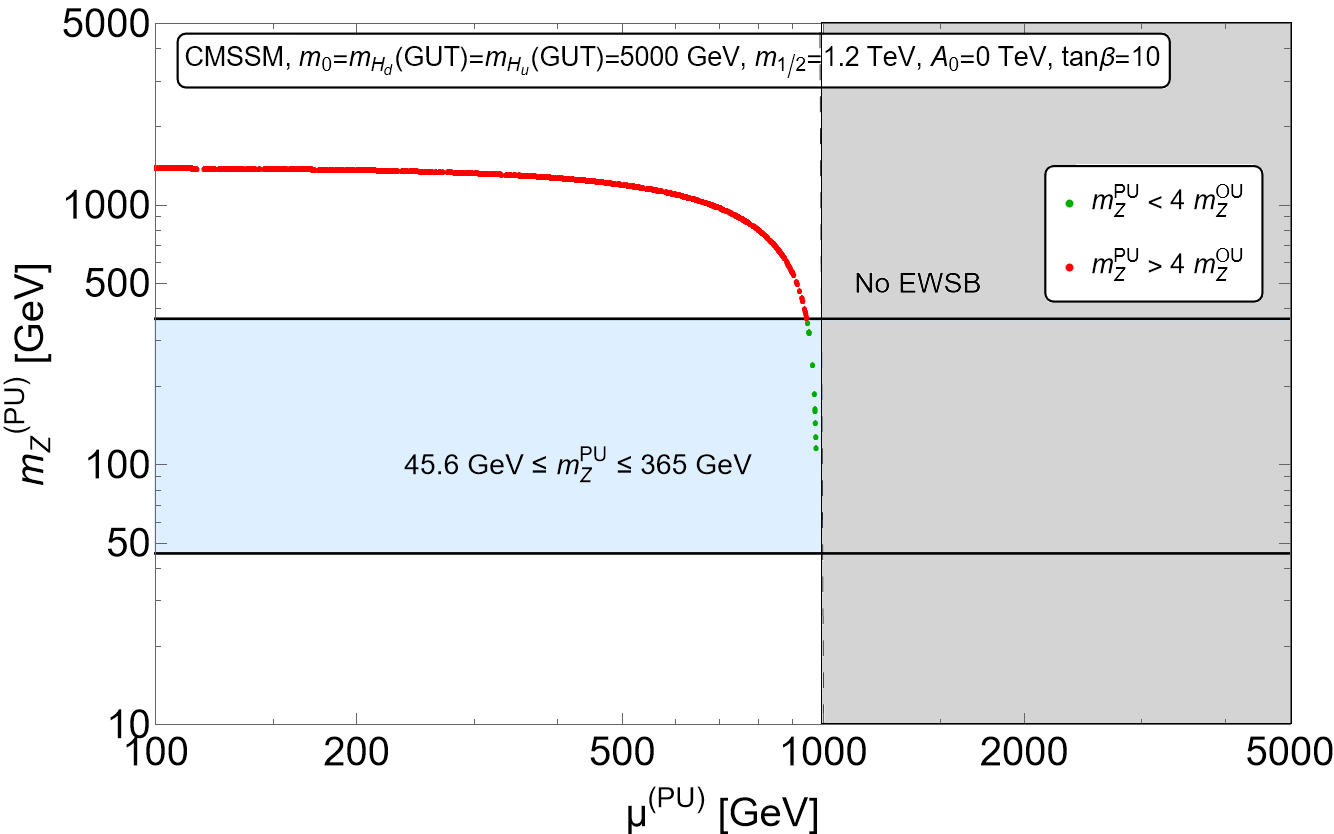}}
  \caption{The value of $m_{weak}^{PU}$ vs. $\mu^{PU}$ 
The green points denote vacua with appropriate
EWSB and with $m_{weak}^{PU}<4 m_{weak}^{OU}$ so that the atomic principle 
is satisfied. Red points have $m_{weak}^{PU}>4 m_{weak}^{OU}$.
}
\label{fig:mweak_muPU}
\end{figure}

We can show this in a different way in this subsection.
In Fig. \ref{fig:mweak_muPU}{\it a}), we adopt as an example our natural
SUSY benchmark model as before, but with variable $\mu^{PU}$.
We then plot the value of
$m_Z^{PU}$ as obtained with our natural SUSY parameter choice but with
varying $\mu^{PU}$. Recall, $\mu^{PU}$ is distributed uniformly across
the decades of values using $f_\mu \sim 1/\mu^{PU}$. From frame {\it a}), we
see a rather large window of $\mu^{PU}$ values from $100-210$ GeV which gives
values of $m_Z^{PU}$ within the ABDS window (green portion of curve).
For larger values of $\mu^{PU}$, $m_Z^{PU}$ drops below the ABDS window
lower bound and then hits the boundary where electroweak symmetry is
not properly broken.

In contrast, in Fig. \ref{fig:mweak_muPU}{\it b}) we instead adopt a value
of $m_{H_u}(GUT)< m_0$ so that $m_{H_u}^2$ is driven to large
(unnatural) negative values at the weak scale. In this case, when we plot
the $\mu^{PU}$ values needed to gain $m_Z^{PU}$ within the ABDS window,
we find a tiny range of parameters around $\mu^{PU}\sim 4$ TeV which is
anthropically allowed. Thus, compared to frame {\it a}), we see that,
given a uniform distribution of $\mu$ parameter on the landscape,
the unnatural model is logically possible-- but highly improbable--
compared to natural SUSY models.

Another example of a finetuned model occurring on the landscape comes
from the CMSSM/mSUGRA model (frame {\it c})) where {\it all} scalar masses are
unified to $m_0$. In this case, with $m_0=5$ TeV, $m_{1/2}=1.2$ TeV,
$A_0=0$ and $\tan\beta =10$ with $\mu >0$, one can see that a
very thin range of $\mu^{PU}$ values around $2$ TeV allow for $m_Z^{PU}$
lying within the ABDS window. Thus, we would expect the CMSSM to also be
rare on the landscape as compared to natural SUSY models which instead
have non-universal scalar masses with $m_{H_u}(GUT)\sim 1.3 m_0$ so that
$m_{H_u}^2$ runs barley negative at the weak scale.

\section{Conclusions}
\label{sec:conclude}

Weak scale supersymmetry (WSS) provides a well-known solution to the gauge
hierarchy problem of the Standard Model via cancellation of all
quadratic divergences in scalar field masses.
WSS is also suppported by a variety of virtual effects, especially gauge
coupling unification and the numerical value of the Higgs boson mass.
$N=1$ SUSY is expected as a byproduct of string theory compactified on a
Calabi-Yau manifold.
In models of string flux compactification, enormous numbers of
vacuum states are possible which allows for Weinberg's anthropic solution
to the CC problem. The question then arises: what sort of soft terms arise
statistically on the landscape, and what are the landscape predictions for WSS?

We addressed this question here via construction of a toy landscape model
wherein the low energy EFT below the string scale was the MSSM but where
each vacuum solution contained different soft term and $\mu$ term values.
Under such conditions, it is expected there is a power-law draw to large
soft terms favoring models with high scale SUSY breaking. However, since the
soft terms and SUSY $\mu$ parameter determine the scale of EWSB, then
anthropics provides an upper limit on the various soft terms: if they lead
to too large a value of the pocket-universe weak scale,
$m_{weak}^{PU}\agt (2-5)m_{weak}^{OU}$ (the ABDS window), then complex nuclei
and chemistry that seems necessary for life would not arise.
This scenario has been used to motivate {\it unnatural} models like the
SM valid to some high scale $Q\gg m_{weak}$, or other unnatural models of
supersymmetry such as split, minisplit, PeV, spread SUSY or high scale SUSY.

Our toy simulation gives a counterexample in that models with
low EW finetuning (radiatively-driven naturalness)
have a comparatively large hypercube of parameter values
on the landscape leading to a livable universe.
For finetuned models, then the hypercube of anthropically-allowed
parameters shrinks to a tiny volume relative to natural models.
Thus, for a landscape populated with vacua
including the MSSM as the low energy EFT, the unnatural models,
while logically possible, are expected to be selected with much
lower probability compared to natural SUSY models characterized by
low $\Delta_{EW}$. This result is simply a byproduct of the finetuning needed
for unnatural models which shrinks their hypercube of allowed parameter
space to tiny volumes. As a result, we expect weak scale SUSY to ultimately
emerge at sufficiently high energy colliders with $m_h\sim 125$ GeV but
with sparticles typically beyond present LHC search limits. In the natural
models, higgsinos must lie in the $100-350$ GeV range while
top-squarks typically lie within the range $m_{\tst_1}\sim 1-2.5$ TeV
with near maximal mixing (due to the landscape pull to large $A_0$ values)
which easily distinguishes them from
unnatural models which would have either far heavier top squarks
or else top squarks with very low mixing.

{\it Acknowledgements:} 

This material is based upon work supported by the U.S. Department of Energy, 
Office of Science, Office of High Energy Physics under Award Number DE-SC-0009956 and DE-SC-001764.
Some of the computing for this project was performed at the OU
Supercomputing Center for Education and Research (OSCER) at
the University of Oklahoma (OU).


\bibliography{landscape6}
\bibliographystyle{elsarticle-num}

\end{document}